# THERMOELECTRIC FIGURE OF MERIT OF BULK NANOSTRUCTURED COMPOSITES WITH DISTRIBUTED PARAMETERS

*A.A.Snarskii[1], A.K.Sarychev[2], I.V.Bezsudnov[3], A.N.Lagarkov[2]*

[1]National Technical University of Ukraine "Kyiv Polytechnic Institute" 03056, Kyiv, Ukraine,

[2]Institute of Theoretical and Applied Electrodynamics, 125412, Moscow, Russia,

[3]Closed JSC "Nauka-Service", 103473, Moscow, Russia

Submitted to editorial board

We consider the effective properties of composite structure that includes nanocontacts between macrocrystallits of bulk material. The model of a such nanostructured composite is developed. In the mean-field approximation we calculate effective values of the thermopower, heat - and the electrical conductivity, thermoelectric figure of merit.

## 1. INTRODUCTION

Developing of a new generation of thermoelectric materials is the target for many research groups. Review [1] contains the current level of research in this very important fundamental and applied area of solid state physics. From a practical point of view, the most important parameter for thermoelectric material is thermoelectric figure of merit $Z$ (figure of merit, Ioffe value). $ZT$ is the dimensionless ratio

$$ZT = \frac{\sigma \alpha^2}{\kappa} T, \qquad (1)$$

where $\sigma$ and $\kappa$ are the electric and thermal conductivities, and $\alpha$ is differential thermopower (the Seebeck coefficient).

Many years of efforts to increase $ZT$ have not yet led to a fundamental breakthrough. (see a review [1] and references in it). Thus, for devices operating at room temperature ($T \approx 300K$), the best material has a value $ZT \approx 1$, as well as a few decades ago. There are published data (reliability of which is unclear) for laboratory samples with $ZT \approx 2-3$ at room temperature. However, commercially available materials with $ZT(T=300K)$ above unity does not currently exist.

There is quite an unusual situation: on the one hand, it has not yet developed the materials with $ZT > 2$ at room temperature, on the other hand, no fundamental limitations is known for the value of $ZT$. Moreover, in a series of publications of G. Casati and co-workers (see [2] and references in it) proposed a "dynamic non-linear machine" - an illustrative model based on classical mechanics, deterministic and stochastic thermostats with their temperatures and electrochemical potentials, which can have $ZT$ arbitrarily high. Of course, such a mechanistic model does not reflect the real situation, but gives us some optimism. In practice, a very limited number of materials are used, only those that can provide both necessary thermoelectric and mechanical and other properties. We believe that improvement of thermoelectric figure of merit is associated now primarily with new phenomena that occur in composite materials.

Nanostructured thermoelectric materials (see, for example, [1,3-7]), such as materials with a superlattice, systems with quantum wells and dots, quantum wires and nanocomposites promise for high $ZT$. In all of these structures increase of $ZT$ can be achieved because of decrease of thermal conductivity due to phonon scattering by inhomogeneities in nanoscale.

Since $ZT$ depends linearly on the conductivity $\kappa$ and quadratic on Seebeck coefficient $\alpha$, a significant increase in the latter allow for the long-awaited breakthrough in the development of thermoelectric devices. One possible implementation involves developing of composite materials with tunnel junctions (TJ) between particles or fibers of bulk matherial. To our knowledge, in [12] the kinetic coefficients of nanostructured thermoelectric materials were first calculated under the assumption that the main role in transport properties of composite plays quantum tunneling between nanoparticles and the background thermal conductivity through TJ is absent. Thermoelectric efficiency of this structure is quite large and, according to [12], $ZT$ can reach values of 2.5 - 4. In this paper we build a model of a composite material with thermoelectric properties originated by to the tunnel junction between particles.

In [12] a model of thermoelectric composite was considered as a chain of connected in series TJ. The model for real composites with nanoscale elements on the surface of bulk semiconductor material must be more complex, it has to take into account the properties of the bulk phase of the particle material, the possibility of a parallel connection of TJ etc. This complex model should show how the composite response to an applied electric and thermal fields, including periodic or changing fields [8], to take into account their mutual influence.

Without theoretical guidance for these plots can not provide reasonable recommendations for the developing of nanostructured thermoelectric converters and devices.

In this paper authors investigate the behavior of effective kinetic coefficients of nanocomposites with electron transport dominated by tunneling of electrons. The mean field theory approach used to find effective conductivity, thermal conductivity and Seebeck coefficient. As a result, the thermoelectric figure of merit is calculated for nanostructured composites and discussed in detail its dependence on the parameters of the composite.

The following section gives a brief review of the properties of single TJ, in Section 3 on the basis of these results a model of a thermoelectric composite with a number of TJ with a wide range of properties is created. In Section 4, this model is used for composites with large $ZT$ values. The Appendix compares the two approaches to the problem posed in sections 3 and 4 percolation-like approach and the mean-field theory and proves the possibility to use the latter.

## 2. THERMOELECTRIC PROPERTIES OF A SINGLE TUNNEL JUNCTION

Thermoelectric properties of TJ are dealt with in many publications. In [9,10] Landauer barriers approach is used, see Fig.1.

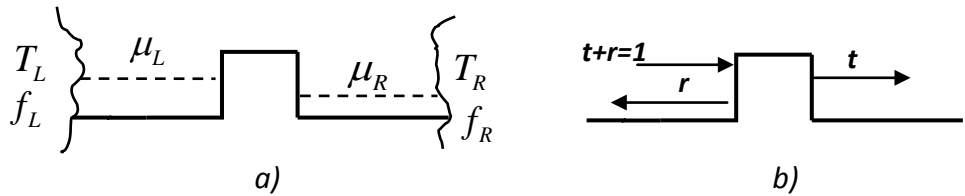

a)                                                                 b)

**Fig.1** Landauer barrier

a. $T_L, T_R$ are temperatures, $\mu_L, \mu_R$ are electrochemical potentials, $f_L, f_R$ are functions of electron distribution in energy (Fermi-Dirac), where indices $L$ and $R$ denote the left and right reservoirs, respectively; b. $t$ is transmission factor, $r$ is factor of reflection from the barrier, $t + r = 1$.

Such a barrier in one-dimensional case represents two reservoirs – "left" and "right" interconnected by a barrier which is described by transmission factor $t$ and reflection factor $r$, so that $r = 1 - t$. According to [10] different effective kine3tic coeffitients are introduced for such Landauer barrier i.e. the Seebeck coefficient $\alpha$, electric conductivity $\sigma$ and thermal conductivity $\kappa$. They merge the barrier properties - $\alpha_b$, $\sigma_b$ and $\kappa_b$ and the contact properties. In the lowest Sommerfeld approximation these kinetic coefficients are expressed via the transmission factor - $t$

$$\sigma_b \sim \frac{t}{1-t}, \quad \kappa_b \sim \frac{t}{1-t}, \quad \alpha_b \sim \left(\ln\frac{t}{1-t}\right)', \qquad (2)$$

where $(...)'$ is electron energy derivative taken on the Fermi level.

For coefficients of the whole Landauer barrier including regard to contacts

$$\sigma_b \sim t, \quad \kappa_b \sim t, \quad \alpha \sim (\ln t)'. \qquad (3)$$

The transmission factor $t$ exponentially strongly depends on barrier thickness, therefore, a slight change in its thickness changes the electrical and thermal conductivities significally (by orders of magnitude), whereby thermopower changes only marginally.

Paper [11] describes experimental study of nanostructured $Sb_{(2-x)}Bi_xTe_3$ that has defects in the form of nanoplates of thickness $\sim 15-20 nm$. It is shown that in such materials $ZT$ increases by 15% as compared to "standard" bulk material (devoid of the above defects).

In paper [12], the kinetic TJ coefficients are calculated and estimation of the thermoelectric parameters of nanostructured $Bi_2Te_3$ is made on the assumption that the main role in charge and heat transport is played by tunnelling of electrons between semiconductor particles. Like in [9,10], the electric and thermal conductivities a strongly (exponentially) depend on barrier thickness, whereas thermopower of such material depends on the barrier thickness weakly (almost linearly).

One should note that thermoelectric figure of merit found in [12] is theoretical estimation and calculated $ZT$ values for TJ are rather higher than actual $ZT$ values but even in this case the qualitative results in [12] will not change either.

Hereinafter, we use as the numerical estimates of parameters of thermoelectric media the properties of the bulk material $Bi_2Te_3$ and the TJ properties calculated in Ref.[12].

# 3. THERMOELECTRIC FIGURE OF MERIT AND THE EFFECTIVE KINETIC COEFFICIENTS OF COMPOSITE WITH NANOCONTACTS

Let us consider thermoelectric composite formed of semiconductor granules with typical size of the order of a micrometer in a matrix of poorly conducting material. In some places the interlayer thickness between granules becomes of the order of a nanometer, see Fig.2.

In such media current carriers have to pass mandatory through places with tunnel junctions in between composite particles. Therefore, to describe the thermoelectric properties of a composite in general, i.e. calculate the effective kinetic coefficients, one should first determine the properties of individual TJ in composite.

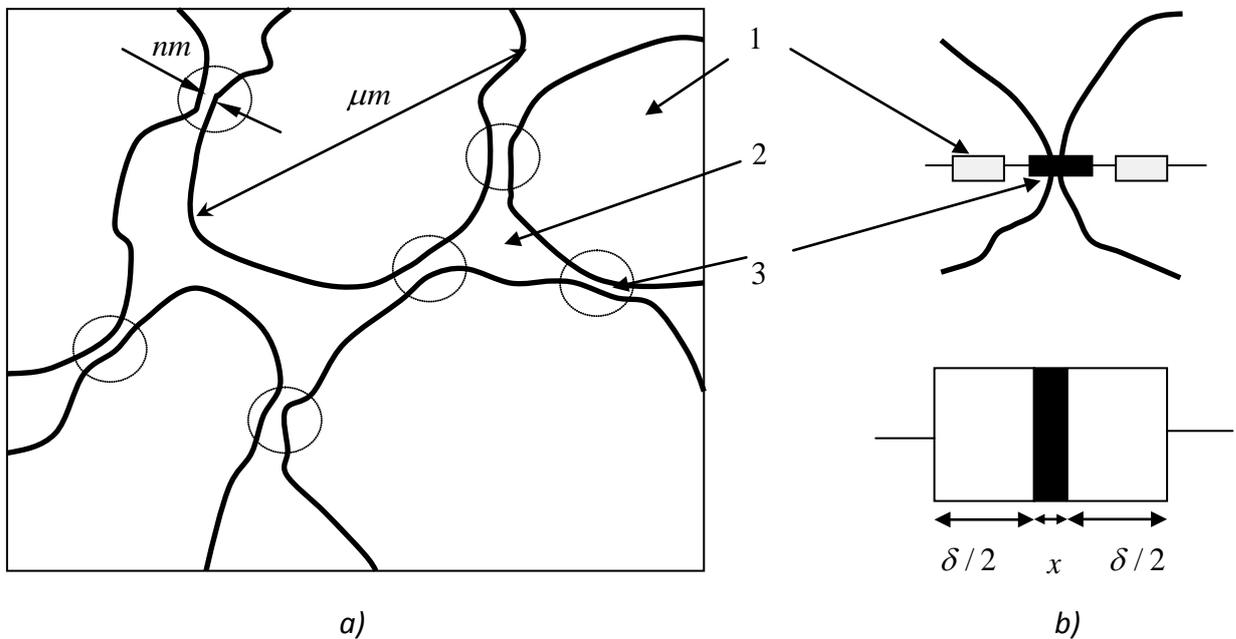

*a)*        *b)*

**Fig.2** Thermoelectric composite structure.

a. Plot of composite geometrical structure. 1- granules of semiconductor (for instance, $Bi_2Te_3$) with typical size $\mu m$, 2- matrix of poorly conducting material, 3 – points of TJ with typical thickness of the order of $nm$. b – electric circuit. 3 – tunnel junction with definite thermopower, the electric and thermal conductivities $\alpha_b$, $\sigma_b$ and $\kappa_b$, below is the plot of single element describing the bulk properties with $\alpha_M$, $\sigma_M$ and $\kappa_M$.

Let us use the results in [12] where thermoelectric figure of merit of single TJ is estimated as $ZT \sim 4$, indeed, it is quite high. Dependence for TJ conductivity $\sigma_b$ on contact thickness $x$ obtained in [12] can be approximated in exponential form with a good accuracy

$$\sigma_b(x) = \sigma_1 e^{-\frac{x-d_1}{\Delta_1}\ln\frac{\sigma_1}{\sigma_2}}, \qquad (4)$$

and the expressions for the TJ Lorentz value - $L_b$ and thermopower - $\alpha_b$ are linear

$$L_b(x) = L_1 + \Delta L \frac{x-d_1}{\Delta_2}, \qquad (5)$$

$$\alpha_b(x) = \alpha_1\left(1 + \frac{\Delta\alpha}{\alpha_1}\frac{x-d_1}{\Delta_2}\right). \qquad (6)$$

Let us set $\varepsilon_b = 0.8 eV$ for the TJ barrier height (one of the two variants given in [12]) and for parameters in (4-6): $d_1 = 0.55 nm$, $\Delta_1 = 0.75 nm$, $\Delta_2 = 1,95 nm$ and $\sigma_1 = 10 \cdot 1/\Omega m$, $\sigma_2 = 0.1 \cdot 1/\Omega m$, $L_1 = 3 \cdot (k_B/e_0)^2$, $\Delta L = 2 \cdot (k_B/e_0)^2$, $\alpha_1 = 300 \mu V/K$, $\Delta\alpha = 60 \mu V/K$ and $(k_B/e_0)^2 = 0.74 \cdot 10^{-8}$.

TJ thermal conductivity $\kappa_b$ depends on the contact thickness and Lorentz value in form $\kappa_b = TL_b\sigma_b$ and can be rewritten as

$$\kappa_b(x) = \kappa_0\left(1 + \frac{L_3 - L_1}{L_1}\frac{x-d_1}{\Delta_2}\right)e^{-\frac{x-d_1}{\Delta_1}\ln\frac{\sigma_1}{\sigma_2}}, \qquad (7)$$

where $\kappa_0 = TL_1\sigma_1$ ($\kappa_0(T=300K) = 0.68\cdot 10^{-4} W/mK$), $(L_3 - L_1)/L_1 = 2/3$.

Parameters of semiconductor bulk material granule (1 – in Fig.2) were assumed equal to those for $Bi_2Te_3$ at $T = 300K$ - $\sigma = 8.3\cdot 10^4 \cdot 1/\Omega m$, $\kappa = 1\cdot W/mK$, $\alpha = 200 \mu V/K$.

According to the proposed model (Fig.2б), current in the composite passes through "packages" representing two parts of the bulk semiconductor material $Bi_2Te_3$ (white colour), divided by TJ (black colour). Having in mind reasonable TJ thickness values $x_{min} = 0.5 nm$ and the fact that conductivity of such TJ is much less than conductivity of the bulk $Bi_2Te_3$, one can consider that the composite resistance is gained on the above "packages" (Fig.2б). Properties of each of these "packages" (three-layer structure where current flow perpendicular to layers) can be found using a well-known solution for thermoelectric properties of flat-layered media [13]

$$\sigma_M(x) = \frac{x+\delta}{\dfrac{x}{\sigma_b(x)} + \dfrac{\delta}{\sigma} + \dfrac{(\alpha_b(x)-\alpha)^2}{\delta \cdot \kappa_b(x) + x\kappa}\delta \cdot x \cdot T}, \tag{8}$$

$$\kappa_M(x) = \frac{x+\delta}{\dfrac{x}{\kappa_b(x)} + \dfrac{\delta}{\kappa}}, \quad \alpha_M(x) = \frac{\dfrac{\alpha_b(x)}{\kappa_b(x)}x + \dfrac{\alpha}{\kappa}\delta}{\dfrac{x}{\kappa_b(x)} + \dfrac{\delta}{\kappa}}. \tag{9}$$

Thus, a composite can be represented as a random lattice where the properties of each bond are determined according to (8-9).

In practice, TJ thicknesses are distributed randomly. For simplicity, we assume that TJ thickness $D(x)$ is uniformly distributed in the range from $x_{min}$ to $x_{max}$, i.e. $D(x_{min} \le x \le x_{max}) = 1$. For some reasonable values for $x_{min} = 0.55 \div 0.75 nm$ and $x_{max} = 1.3 nm$ the thickness of TJ changes insignificantly, not more than a factor of three. In opposite, TJ conductivity $\sigma_b$ changes by orders of magnitude, and "package" conductivity $\sigma_M(x)$ changes just as much. Thus, current and heat flux distribution is strongly inhomogeneous. A lattice with such (exponentially wide) distribution of conductivities is called Miller-Abrahams lattice, its properties are studied in a large number of papers, see, for instance, [14,15]. In the Appendix, two methods for solving the problem of the effective conductivity in such network are described and it is shown that for qualitative estimates in the case under study one can use the average field theory approximation.

The simplest method of average field approximation for thermoelectric effects in two-phase media proposed in [16], see also [17] and can be written as

$$\left\langle \frac{\sigma_e - \sigma}{2\sigma_e + \sigma} \right\rangle = 0, \quad \left\langle \frac{\kappa_e - \kappa}{2\kappa_e + \kappa} \right\rangle = 0, \tag{10}$$

and

$$\alpha_e = \frac{\langle \alpha\sigma/\Delta_0 \rangle}{\langle \sigma/\Delta_0 \rangle}, \tag{11}$$

where $\langle ... \rangle$ means sum over phases and $\Delta_0 = (2\sigma_e + \sigma)(2\kappa_e + \kappa)$.

In this case, phases are numbered by a continuous parameter – TJ thickness $x$, namely $\sigma \equiv \sigma_M(x)$, $\kappa \equiv \kappa_M(x)$, $\alpha \equiv \alpha_M(x)$, and under $\langle ... \rangle$ we understand parameter averaging, i.e.

$\langle f(x) \rangle = \int_{x_{\min}}^{x_{\max}} f(x)D(x)dx$. Thus, a system of equations for the calculation of the effective electric and thermal conductivity represents two nonlinear integral equations

$$\int_{x_{\min}}^{x_{\max}} \frac{\sigma_e - \sigma_M(x)}{2\sigma_e + \sigma_M(x)} D(x)dx = 0, \quad \int_{x_{\min}}^{x_{\max}} \frac{\kappa_e - \kappa_M(x)}{2\kappa_e + \kappa_M(x)} D(x)dx = 0, \tag{12}$$

and the effective thermopower is determined as follows

$$\alpha_e = \frac{\int_{x_{\min}}^{x_{\max}} \frac{\alpha_M(x)\sigma_M(x)}{\Delta_0(x)} D(x)dx}{\int_{x_{\min}}^{x_{\max}} \frac{\sigma_M(x)}{\Delta_0(x)} D(x)dx}, \tag{13}$$

where $\Delta_0(x) = (2\sigma_e + \sigma_M(x))(2\kappa_e + \kappa_M(x))$.

Fig.3 shows behavior of the effective conductivity $\sigma_e$ and thermal conductivity $\kappa_e$ and Fig. 4 shows behavior of the Seebeck coefficient and figure of merit.

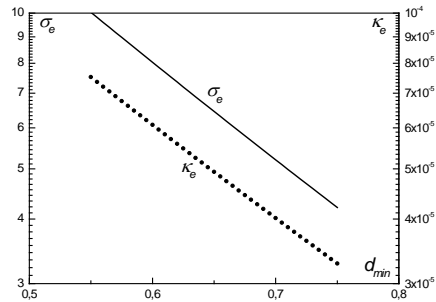

**Fig.3** Dependences of the effective conductivity $\sigma_e$ and thermal conductivity $\kappa_e$ on the TJ $d_{\min}$.

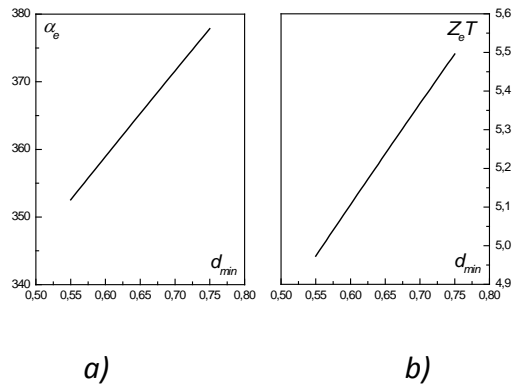

*a)* *b)*

**Fig.4** Dependence a) of the effective Seebeck coefficient (in units $\mu V / K$) b) of the effective coefficient of thermoelectric figure of merit $Z_e T$ on the value of TJ minimal thickness $d_{\min}$.

The effective Seebeck coefficient shows a weak, almost linear dependence on the minimal thickness of TJ (Fig.4a) existed in a composite, unlike the exponentially strong dependence of the electric and thermal conductivity of TJ (4), (7) and "package" (8), (9) upon thickness of TJ. The same is true for behaviour of thermoelectric figure of merit (Fig. 4b), that can be calculated as $Z_e T = \sigma_e (\alpha_e)^2 T / \kappa_e$,

## 4. THERMOELECTRIC FIGURE OF MERIT AND THE EFFECTIVE KINETIC COEFFICIENTS OF A COMPOSITE WITH HIGH $ZT$

A more complicated version of average field theory approach in [17] allows to take into account high $ZT$ values ($ZT > 1$). According to [17], the effective coefficients $\sigma_e, \kappa_e$ and $\alpha_e$ can be found by solving a system of three nonlinear integral equations

$$\int_{x_{\min}}^{x_{\max}} A(x) D(x) dx = 1, \quad \int_{x_{\min}}^{x_{\max}} B(x) D(x) dx = 0, \quad \int_{x_{\min}}^{x_{\max}} C(x) D(x) dx = 0 \tag{14}$$

where

$$A(x) = 3 \frac{\sigma_e (2\chi_e - \chi_M(x)) - \gamma_e (2\gamma_e - \gamma_M(x))}{\Delta(x)},$$

$$B(x) = 3 \frac{\gamma_e \chi_M(x) - \gamma_M(x) \chi_e}{\Delta(x)}, \tag{15}$$

$$C(x) = 3 \frac{\gamma_e \sigma_M(x) - \gamma_M(x) \sigma_e}{\Delta(x)}.$$

Here

$$\gamma_M(x) = \sigma_M(x) \alpha_M(x), \quad \chi_M(x) = \frac{\kappa_M(x)}{T} + \sigma_M(x) \alpha_M^2(x),$$

$$\gamma_e = \sigma_e \alpha_e, \quad \chi_e = \frac{\kappa_e}{T} + \sigma_e \alpha_e^2, \tag{16}$$

$$\Delta(x) = (2\sigma_e + \sigma_M(x))(2\chi_e + \chi_M(x)) - (2\gamma_e + \gamma_M(x))^2$$

This system is solved numerically for $d_{\min} = 0.55 nm$, $T = 300K$ (values of other parameters are given next to (6)).

The effective figure of merit with a given set of parameter values is equal to

$$Z_e T = \frac{\sigma_e \alpha_e^2}{\kappa_e} T = 3.52. \tag{17}$$

This figure of merit value is somewhat lower than the minimal figure of merit for that of tunnel junction

$$Z_eT\big|_{\min} = 3.67. \qquad (18)$$

Thus, a considerable resistance distribution in a composite and, hence, a considerable inhomogeneity of current density in it do not affect and do not reduce the figure of merit.

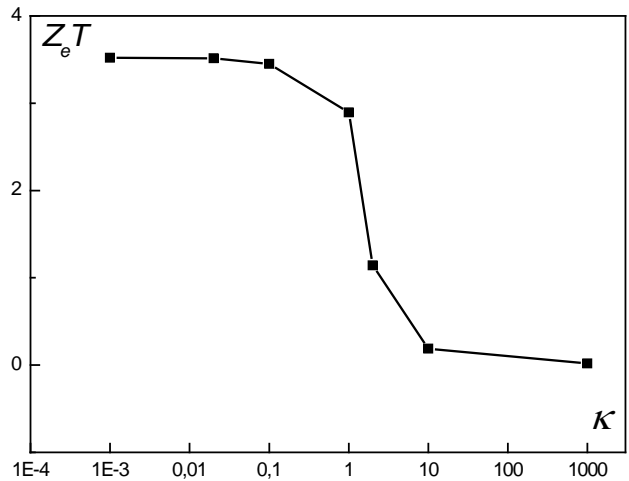

**Fig.5.** The effective thermoelectric figure of merit versus the thermal conductivity of the bulk phase.

It is interesting and important to note that the figure of merit of composite $Z_eT$, (see Fig.5), is monotone (as it should be) depends on the thermal conductivity of the bulk phase, i.e. increases monotonically with its reduction. The unexpected result is the fact that this dependence in the range from $0.1 \div 10 W/m \cdot K$ experiences a drastic drop, remaining nearly constant beyond it.

## 5. CONCLUSION

Methods of average field theory were used to make a theoretical description of the behavior of a thermoelectric composite consisting of macroparticles but including nanoelements – tunnel junctions. With regard to numerical values for model parameters obtained earlier in [12] developed model of thermoelectric composite proves possibility to calculate the thermoelectric figure of merit of such composite. Calculated estimations allow to look with optimism to the possibility of obtaining high $Z_e T$ values in materials that are not traditionally qualified as nanomaterials. The bulk granules of macroscopic size here are "carriers" of nanoelements on its surface - TJ with high thermoelectric figure of merit, at the same time "carriers" connect these TJ nanoelements in such way that thermoelectric parameters of a composite prove to be comparable to properties of TJ themselves. Therefore, composites with TJ of investigated structure are theoretically promising creation of high-performance thermoelectric materials.

Certainly, the proposed composite structure wherein nanoelements are of key importance on passing of current and heat flux is not the unique one. Thus, for instance, of great interest are composites including the so-called nanowires [18] and many other structures.

Of course, the proposed structure of the composite, where nanoelements are key elements for current and heat flow is not unique. For example, great interest is a composite, comprising the so-called nanowires [18] and many other structures.

**APPENDIX 1**

Consider a problem of the effective conductivity $\sigma_e$ on the Miller-Abrahams lattice, with the local conductivity $\sigma(\xi)$

$$\sigma(\xi) = \sigma_0 e^{-\lambda \xi}, \qquad (A1)$$

where random variable $\xi \in [0,1]$ is distributed uniformly, and parameter $\lambda$ means the degree of medium inhomogeneity $\lambda \gg 1$.

The percolation-like approach to find $\sigma_e$ is based on the following idea. All lattice resistances (bonds) are temporarily "taken out" from their places, and then in turn they are "inserted" in value ascending order. The first resistance that close the lattice (create the so-called infinite cluster) will specify the resistance of the entire lattice, since it is the largest in this cluster. All the resistances connected afterwards will be considerably larger ($\lambda \gg 1$) and unable to perform the function of a bypass. With such an approach

$$\sigma_e = \sigma(\xi) = \sigma_0 e^{-\lambda \xi_c}, \qquad (A2)$$

where $\xi_c = p_c$ is a conventional percolation threshold of the two-phase system [14,15].

In many cases the expression (A2) gives reasonably precise description of the effective conductivity of lattice. However, in some cases, in particular, at determination of the correlation length of such lattice, in the expression for the effective conductivity $\sigma_{e(perc)}$ one should take into account the pre-exponential factor, namely

$$\sigma_{e(perc)} = \sigma_0 \lambda^{-y} e^{-\lambda \xi_c}. \qquad (A3)$$

To determine critical index $y$ is a much more complicated problem. According to [19,20], critical index $y$ is expressed through critical indexes of conductivity and for the three-dimensional case is equal to $y = (t-q)/2 = 0.635$.

The average field approach (Bruggeman-Landauer approximation), proposed in [21] and [22] for a two-phase medium is an equation from which one can find the effective conductivity

$$\frac{\sigma_e - \sigma_1}{2\sigma_e + \sigma_1} p + \frac{\sigma_e - \sigma_2}{2\sigma_e + \sigma_2}(1-p) = 0, \tag{A4}$$

where $p$ is concentration of the first phase with conductivity $\sigma_1$.

For $n$ - phases (A4) is whiten as

$$\sum_{k=1}^{n} \frac{\sigma_e - \sigma_k}{2\sigma_e + \sigma_k} p_k = 0, \tag{A5}$$

where $p_k$ is concentration of phase with conductivity $\sigma_k$.

For a continuous distribution of phases (A5) is readily generalized and represents a nonlinear integral equation

$$\int_{\xi_1}^{\xi_2} \frac{\sigma_e - \sigma(\xi)}{2\sigma_e + \sigma(\xi)} D(\xi) d\xi = 0, \tag{A6}$$

where in the case under study (A1) $\xi_1 = 0, \xi_2 = 1, D(\xi) = 1$ and, thus, is of the form

$$\int_0^1 \frac{\sigma_e - \sigma(\xi)}{2\sigma_e + \sigma(\xi)} d\xi = 0. \tag{A7}$$

For the case $\sigma(\xi) = \sigma_0 e^{-\lambda \xi}$ the expression (A6) can be written as follows

$$\int_{\sigma_0}^{\sigma_0 \exp(-\lambda)} \frac{\sigma_e - \sigma}{2\sigma_e + \sigma} f(\sigma) d\sigma = 0. \tag{A8}$$

where $f(\sigma) = 1/\lambda\sigma$ and the integral of (A8) can be found analytically and the expression for effective conductivity is a function of inhomogeneity parameter $\lambda$ [23]. This value of effective conductivity will be denoted as $\sigma_{e(mf)}$.

It is surprising enough that the mean-field approach usually used for not very large heterogeneity and based on solving the problem of the distribution of electric fields and currents around an isolated conducting sphere describes quantitatively the dependence of the effective conductivity on the parameter $\lambda$, i.e. for high inhomogeneity (see Fig. A1).

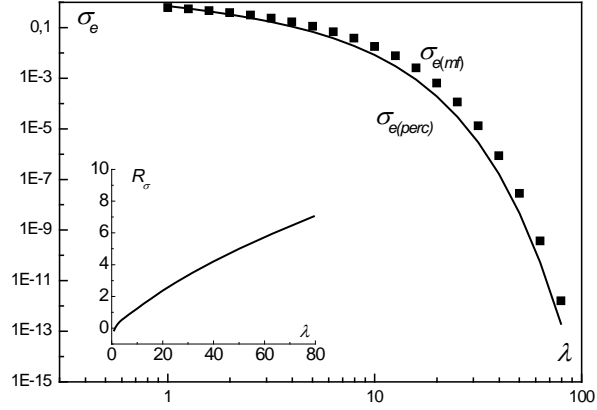

Fig.A1. Dependence of the effective conductivity $\sigma_e$ on the inhomogeneity parameter $\lambda$ in average field approximation $\sigma_{e(mf)}$ and at percolation-like approach – $\sigma_{e(perc)}$; On the inset – relative divergence of the two methods $R_\sigma = (\sigma_{e(mf)} - \sigma_{e(perc)}) / \sigma_{e(perc)}$.

For a large inhomogeneity, for instance, $\lambda \sim 100$, when the ratio of individual network resistances can reach $10^{27}$, the difference in the effective conductivity values calculated by the average field method (A8) and the percolation-like method (A3) is not more than 8 fold.

According to (4), the expression for conductivity $\sigma_b$ can be written as

$$\sigma_b(x) = const \cdot e^{-\frac{x}{\Delta}\ln\frac{\sigma_1}{\sigma_2}}, \qquad (A8)$$

Therefore, if (A8) is re-written in form of (A1), then parameter $\lambda$ will be equal to $\lambda \approx 4.6$. And, thus, to calculate the effective properties of a medium with distributed properties (8) – (9), one can use with a good precision the average field approximation.